\documentclass[5p,authoryear]{elsarticle}

\usepackage[T1]{fontenc}
\usepackage{lmodern}
\usepackage{booktabs}
\usepackage{amsmath,amssymb}
\usepackage{array}
\usepackage{xurl}
\usepackage{xcolor}
\usepackage{tabularx}
\usepackage{eso-pic}
\usepackage{graphicx}
\usepackage{float}
\usepackage{microtype}
\usepackage{tikz}
\usetikzlibrary{positioning}

\makeatletter

\long\def\@makecaption#1#2{%
  \vskip\abovecaptionskip
  {\footnotesize\noindent #1. #2\par}%
  \vskip\belowcaptionskip}
\makeatother

\newcommand{\cfull}{full match}
\newcommand{\cskel}{skeleton match}
\newcommand{\cfollow}{adherence}
\newcommand{\ours}{Conditioned}
\newcommand{\oneshot}{One-stage}
\newcommand{\ce}{Matched-step}
\newcommand{\ew}{Equal-weight}
\newcommand{\goldc}{Oracle-contract}
\newcommand{\renc}{Renamed}
\newcommand{\shuf}{Shuffled}
\newcommand{\pt}{\,pt}

\journal{}
\biboptions{sort}

\begin{document}
\setlength{\emergencystretch}{2em}

\begin{frontmatter}

\title{Construction Contracts for Generated Embedded Domain-Specific Language Programs:\\
A Measurement Protocol and a Conditional Effect}

\author[a]{Haowen Guo\corref{cor1}}
\ead{2025091203013@std.uestc.edu.cn}
\author[a]{Yangzhao Huang}
\ead{yangzhao_se2025@163.com}
\cortext[cor1]{Corresponding author.}
\affiliation[a]{organization={School of Information and Software Engineering,
  University of Electronic Science and Technology of China},
  addressline={No.\ 2006 Xiyuan Avenue, High-Tech Zone (West)},
  city={Chengdu},postcode={611731},state={Sichuan},country={China}}

\begin{abstract}
A generated domain-specific program can be accepted by its host and still miss the
reference construction. We measure the two events separately and test a short
construction contract.

The vehicle is Open Quantum Assembly Language (OpenQASM)~3. One 3-billion-parameter
generator is run under seven conditions on 100 quantum approximate optimization
(QAOA) and 100 variational quantum eigensolver (VQE) instances. Eighty of 100 QAOA
signatures appear in some training split. Exact McNemar tests and 95\% paired
bootstrap intervals are used. The contract, a card, lists structural decisions,
enters the prompt, and is checked without a simulator.

Parse rates are 0.55 to 0.97 and full structural match is 0.02 to 0.87. On QAOA, at
fixed weights, the card moves full match by +32.0 percentage points on card-trained weights
(interval +22 to +42) and by $-13.0$ ($-23$ to $-3$) without that training. The +45.0-point
difference is not a deployment contrast. The same-step plan-free
gain is +23.0 (+13 to +33); the primary distributional score does not separate it.
Against the earlier checkpoint the gain is +27.0 (+17 to +37), and the checkpoint
also changes. A rule card is not separated from the learned planner, nor is a
predicted-card compiler (0.760 vs.\ 0.780). On 20 QAOA rows outside every training split,
the interval includes zero. Under this card, VQE adherence is 0.020 and full match
stays near 0.03.

Report the two checks separately. Insert a card only into a generator trained to
read it, and only where a pilot shows adherence.
\end{abstract}


\end{frontmatter}

\section{Introduction}\label{sec:intro}

Tool chains increasingly delegate a fragment of work to a large language model: a
query, a constraint, a configuration unit, a circuit description. Each is an embedded
domain-specific program. Its text means little on its own and much relative to an
interface the host has already fixed, and the interface decides whether the fragment is
accepted. Acceptance is weaker than correctness, because a fragment that parses can
still violate the construction the host expects: the terms it covers, the symbols it
shares, the order in which it commits to them. A consumer downstream can reject a
fragment the parser accepted.

Software engineering has met this gap before, under other names.
Text-to-SQL reports exact match beside execution accuracy, because the two disagree on
queries that are right in different ways \citep{spider2018}, and the oracle problem says
the same thing more generally: what decides correctness is the criterion the check
applies, not a property of the program alone \citep{barr2015}. Code-synthesis metrics add
a syntax-level term to string similarity for the same reason \citep{ren2020}. What these
leave open is the operational question. Can a reference construction be recovered from
the text by a walk over it, reported field by field, computed without a simulator, and
then used as a design target? That is the middle object this paper measures, and it sits
between the two numbers that evaluations of generated code usually report.

The measurements come from one such interface. A fine-tuned generator writes
parameterized quantum circuits in OpenQASM~3 \citep{cross2022}, and a fixed
host interface decides what it may declare and how many lines it may use. Across the 200
evaluation instances of the main comparison the parser accepts most of what it is given; far less of it matches the reference. In
the most divergent cell 0.810 of the outputs parse and 0.020 match all six structural
fields. The host can accept the artifact even
though the circuit encoded in it differs structurally from the reference.

We then place a specification between the instance and the artifact and ask whether it
moves that second number. We call it a \emph{construction contract} (a \emph{card}): a finite
product of constant-size structural decisions. A walk extracts the card from a reference
artifact, and a forward pass predicts it from the instance text. The word is narrower
than design by contract \citep{meyer1992}. There is no precondition, no postcondition, and no runtime
assertion. The card records a reference construction. We score it as prompt content and
as a text-walk predicate, and the prompt block is its serialized form. Seven conditions
and one relevance control separate slot content from slot presence, from prompt format,
from extra training, and from the weights that read the card. The two families differ in
construction, so the boundary can be measured rather than assumed.

Three findings organize what follows. Parse rate and full structural match move apart
on the same texts. On QAOA the card raises full match only after card-conditioned
training, and that estimate is conditional on the constructed set. On VQE, under this
card, adherence stays near zero and full match does not rise. After a shared refit,
the primary distributional measure does not separate the matched-step control
(Sections~\ref{sec:semantics} and~\ref{sec:construct}). Circuits are the vehicle.
Section~\ref{sec:disc-transfer} states the transfer test, and Section~\ref{sec:threats}
records the limits.

\subsection{Contributions}\label{sec:contrib}

The paper makes three contributions.

\begin{enumerate}
\item A two-predicate measurement protocol for generated domain-specific language (DSL) artifacts. Interface
acceptance (parse) and reference structural conformance (full match) are separate
columns. Skeleton match is the name-quotiented companion, and semantic quality is
reported beside them rather than as a third predicate. The structural predicate is a
walk over the text. It calls no simulator, and it names the field that failed
(Sections~\ref{sec:measures}, \ref{sec:parse} and~\ref{sec:fields-fail}).
\item A checked, corruptible intermediate scored as prompt content and as a structural
predicate, with controls for content, format, extra training, and a planner-free card.
On this QAOA template, card content and card-conditioned training are jointly
necessary. At fixed weights the predicted card is worth $+32.0\pt$ of full match on
card-trained weights and costs $13.0\pt$ on weights that never saw a card. The
$+45.0\pt$ gap summarizes that two-by-two and is not a deployment contrast. A rule
card and a model-free compiler of the predicted card are not separated from the
learned pipeline at this $n$. The pipeline contrast against one-stage also changes
the checkpoint and is reported as a deployment figure
(Sections~\ref{sec:effect}, \ref{sec:attrib} and~\ref{sec:disc-report}).
\item A boundary, read on a second family of the same data rather than as a pre-specified
moderator. Under the card as defined here, VQE adherence is 0.020 and full match stays
between 0.02 and 0.04; the reference card itself leaves full match at 0.030 while breaking
the interface (parse 0.550). That pattern is consistent with an incomplete field list
(no parameter budget) but is not separated from other generation-side failures. The 7B
probe raises adherence to 0.210 on the reference card and to 0.172 on the predicted one,
while full match on the predicted card stays at 0.070; it has no base stage and no
plan-free continuation, so it is reported as an observation
(Section~\ref{sec:vqe}, Table~\ref{tab:7b-vqe}).
\end{enumerate}

The extractor, checker, prompts, per-row generations, and evaluation scripts are
deposited as \citet{guo2026artifact} (Section~Data availability). Every table can be
rebuilt from the per-row logs in that deposit. The raw circuit corpus has no public
identifier and is not redistributed. Third parties can recompute the published tables
from the deposited logs and cannot regenerate the reference circuits from a public
problem set.

The scope is deliberately narrow. No new generator, intermediate language, or claim
about the best way to synthesize circuits is offered. The generator is held fixed
across conditions, because the measurements concern the artifact and the practice. Structural conformance is reference-relative, and
Section~\ref{sec:construct} measures how far it stands from semantic equivalence. The field
list is local to the corpus that produced it, and Section~\ref{sec:disc-transfer} states
what would have to hold for another embedded language.

\section{Background and related work}\label{sec:related}

Pass@$k$ summaries fit general-purpose languages with executable tests
\citep{chen2021,austin2021}. An embedded DSL fragment is weaker for that protocol:
the language is low-resource, ``running it'' may mean checking a relation, and the host
interface that consumes the fragment is not the test harness. Quantum code-generation
benchmarks make the gap concrete---accuracy below 0.40 with failures sorted into API,
construction, and logic errors \citep{guo2025}---but still score the finished program
without separating interface acceptance from construction match
(Section~\ref{sec:parse}).

Software engineering already knows related separations. Insufficient tests can inflate
pass rates by tens of points \citep{liu2023}; DSL tooling is itself an engineering
artifact \citep{ratiu2018}; text-to-SQL and the oracle problem separate exact match from
execution accuracy \citep{spider2018,barr2015}; and constraint-language generation in this
journal separates well-formedness from correctness \citep{delgado2026}. What those
traditions still leave open for a host-embedded fragment is a predicate computable from
the text alone---no suite, no simulator---that recovers the construction field by field.

End-to-end execution feedback is expensive when the event is rare
\citep{runeson2009,wohlin2012,quasar}, and unaided self-repair rarely helps without a
localizing external signal
\citep{madaan2023,chen2023,olausson2024,huang2024,kamoi2024}. That is why the contract sits
before generation, not as a critique of a finished circuit.

Sketches \citep{solar2008}, constrained decoding \citep{picard2021}, free-text
self-plans \citep{jiang2024}, human or protocol-guided plans \citep{agcot,pg}, and
Spider-style component match \citep{spider2018} each stop short of one intermediate
that is predicted from the instance, withheld or corrupted, and scored both as prompt
content and as a structural predicate (Section~\ref{sec:criterion}). A free-text plan
is not a finite field vector. Constrained decoding rejects illegal tokens during
generation. The card is checked afterward and can be legal yet wrong for the
instance (Section~\ref{sec:relevance}). A sketch is a partial program. The card does
not emit gates until an executor or a generator realizes it.

The word \emph{contract} will be heard as design by contract \citep{meyer1992}. Those traditions
constrain behavior at a boundary. The object studied here records a reference
construction and is neither a precondition nor a runtime check
(Section~\ref{sec:reduction}). Host-language embedding and language-workbench
engineering are the natural neighbors of a fragment the parser accepts while the
construction is still wrong. This study runs neither, and multi-file repair is out of
scope. Full match certifies agreement with a fixed construction, which
Table~\ref{tab:assoc} shows is not a near miss on fitted energy
\citep{runeson2009,wohlin2012,ralph2021}. A single scalar often mixes the three
events \citep{evtikhiev2023}. Sections~\ref{sec:measures}
and~\ref{sec:parse} give each event its own column, and Section~\ref{sec:contam}
measures training--evaluation overlap instead of assuming a clean split.

\section{Study design}\label{sec:design}

We compare generation conditions on a fixed set of instances, and we hold the
generator, the decoder, and the checker fixed. Only two things move: the checkpoint,
when a control requires it, and the text placed in the contract slot at inference.

\subsection{Research questions}\label{sec:rq}

Four questions organize the measurements, and for each one Table~\ref{tab:rq}
fixes the comparison that answers it and the measure that comparison may use.

\begin{enumerate}
\item \textbf{Interface and construction.}
Do generated artifacts fail because they do not parse, or because the
construction inside a parsed artifact is wrong? Are the parse rate and the
structural match separable within a single condition? Semantic scores are
taken up in Section~\ref{sec:semantics}.
\item \textbf{Effect of the contract.}
When the generator is given a contract predicted from the instance, does
structural match rise relative to a one-stage generator, and relative to a
generator continued for the same number of steps without a contract?
\item \textbf{What the gain depends on.}
Is a rise in structural match accounted for by extra training, by the
checkpoint, by the presence of a formatted block, or by the content and the
names inside the contract?
\item \textbf{Where the contract is realized.} Does the generator carry the reference
contract out on both families, and does structural match rise where it does?
\end{enumerate}

\begin{table*}[t]
\centering
\caption{Research design. Each question, the comparison that answers it, and the
measure that comparison uses. Supporting measurements of solution quality and
of benchmark overlap are reported with the question they qualify.}
\label{tab:rq}
\scriptsize
\begin{tabularx}{\textwidth}{@{}l >{\raggedright\arraybackslash}X >{\raggedright\arraybackslash}X@{}}
\toprule
Question & Comparison & Measure \\
\midrule
1 & All seven conditions, both families &
  Parse rate against \cfull{} \\
2 & \ours{} against \oneshot{}; \ours{} against \ce{} &
  \cfull{} \\
3 & \ew{}, \ce{}, \renc{}, \shuf{}, the rule card, and A1 &
  \cfull{} and \cskel{} \\
4 & QAOA, where the generator carries the reference contract out, against VQE, where the 3B generator does not &
  \cfull{} and \cfollow{} \\
\bottomrule
\end{tabularx}
\end{table*}

\subsection{Instances and artifacts}\label{sec:instances}

An artifact is a parameterized circuit in OpenQASM~3, written against a fixed host
interface: one declaration \texttt{input float[64] x}, no ancilla qubits, no
mid-circuit measurement, and a line budget set by the family. This is the object a
tool chain accepts or rejects, and it is small enough for a structural checker to
read without a human judge and without a simulator.

Two families appear because their constructions differ. That difference is what makes question~4 decidable. The variational family is not a second copy of the sample: it is where adherence collapses under the card as defined in this study, so the design can show a boundary rather than assume one. Quantum approximate optimization (QAOA) alternates cost and mixer layers on a problem graph \citep{farhi2014}. The variational quantum eigensolver (VQE) repeats a motif drawn from an ansatz family \citep{peruzzo2014}.
The evaluation set holds 200 instances, 100 per family, and it is not a uniform sample of the test split. A fixed rule first keeps every row whose signature is absent from the QAOA and VQE training splits, with no further sampling. That family-clean cell (Table~\ref{tab:overlap}) has 46 QAOA rows and 82 VQE rows, 128 in all. Each family is then filled to 100 from rows whose signature was seen, stratified by qubit count ($\le 8$, $9$--$12$, $>12$) at a fixed seed, so these absent rows are over-represented. Of the 128 family-clean rows, 102 are absent from every split, against the 17 such rows the parent split would contribute. Of the 46 family-clean QAOA rows, 20 are absent from every split; all 82 family-clean VQE rows are. Across the 200 rows the qubit bands are 75, 21 and 4 for QAOA and 17, 65 and 18 for VQE, and depth runs from $p=1$ to $4$ (62, 12, 15 and 11 QAOA instances; 64, 10, 14 and 12 for VQE). The qubit counts are uneven by construction.

Each instance carries the problem type, the graph or the ansatz, the cost operator,
the number of qubits, and, for QAOA, the depth; it carries no energy, no optimum and
no reference parameter vector. Recovering the construction from that text is the
generator's task.

A separate corpus of 10{,}207 reference circuits serves only to derive and audit the
contract (Section~\ref{sec:contract}); inference never sees it.

\subsection{Measures}\label{sec:measures}

Table~\ref{tab:measures} fixes the measures. Two of them are easy to conflate,
and the rest of the paper treats them as different predicates.

\emph{Reference-based structural conformance}, abbreviated \emph{full match} in the
tables, is true when every structural field of the reference construction agrees
with the emitted artifact. QAOA has five fields:
the covered terms, the coefficient slot of each symbol, the allocation of
symbols, the order of lines, and inversion. Order and cover coincide on this corpus,
because the reference order is the cover order, so the conjunction has four independent
fields on QAOA and order is carried as a check.
VQE has six: the motif, the
repetition count, the sharing pattern, the tail, the gate set, and the arity.
\emph{Skeleton match} quotients symbol names and compares only the
cover and the slot sequence. It is defined for QAOA. On VQE it is undefined,
and the tables leave the cell blank. A blank cell is not a failure. The
difference between the two predicates mixes field strictness with naming, so a
naming effect is read from \cskel{} and a full-field effect from \cfull{}.

The measure certifies agreement with a construction a tool chain has fixed, not
correctness in the abstract. Three layers stay apart: interface validity (the text
parses and declares exactly the parameters it uses), reference structural
conformance, and semantic quality. Only the middle layer is the primary measure.
Conformance is a walk over the text, so it is sufficient for mechanical audit. It is
not necessary for semantic equivalence. An artifact can implement the same state
with a different construction (Section~\ref{sec:construct}). It does not replace
semantic validation, which is why those measures are reported beside it. A tool that
reads the construction would make the predicate an engineering property. This study
runs no such tool (Section~\ref{sec:th-construct}).

\emph{Adherence} is true when the artifact implements the contract
it was given, whether or not that contract matches the reference. It is
computed for the conditioned condition only. Full match remains the primary
structural measure, because a generator can adhere to a wrong contract.

Scores on the parameters as emitted bind every symbol parameter to the constant 0.1, and
the reference circuit the same way; the 0.1 here is not the KL threshold below. That protocol
is degenerate (Section~\ref{sec:parse}). The semantic measures are therefore computed from a refit of the reference circuit and of every emitted circuit under one budget:
COBYLA, five restarts, 250 iterations, seed 0 on each call. \emph{HQCR} is true when the
Kullback--Leibler divergence of the reference output distribution from the emitted one is at
most 0.1. It is the primary semantic measure because it does not saturate: the one-sided
energy hit reaches 1.000 on both arms of the reference-contract condition and cannot separate
them.

Divergence is measured over computational-basis outcomes. Both circuits of a pair have
the same number of qubits, so the supports coincide. Each probability is floored at
$10^{-12}$ and the vectors are renormalized, with the reference on the left of
$D_{\mathrm{KL}}(p\,\|\,q)$. The reported quantity is this floor-regularized divergence.
Recomputing the units whose divergence falls between 0.02 and 0.6 at floors of
$10^{-6}$, $10^{-9}$ and $10^{-12}$ moves the largest value by $6\times10^{-3}$, and no
verdict changes. By Pinsker's inequality, a divergence of 0.1 keeps total variation at
or below about 0.22 on the regularized quantity. The contrast is stable in the cut
($+8.9$ points at 0.1, $+8.9$ at 0.05, and $+11.1$ at 0.3;
Section~\ref{sec:semantics}). The one-sided hit at tolerance $\tau$, written
$q$-hit@$\tau$, is true when the fitted energy is at most the reference energy plus
$\tau$. The two-sided score is true when the absolute deviation is at most 0.2. VQE
is not given a semantic conclusion, because adherence there is too low for the
contract to have been executed. The arms emit a median of 93 parameters against 42 in
the reference circuits. Splitting the variational rows at that median does not move
the outcome (small half SREV 0.108 and HQCR 0.027; large half 0.094 and 0.031).

\begin{table*}[t]
\centering
\caption{Measures. Interface and structural measures are
functions of the emitted text. Semantic measures are computed after a shared
parameter refit and are interpreted on QAOA, where that refit is closed.
Full match and skeleton match are different predicates.}
\label{tab:measures}
\scriptsize
\begin{tabularx}{\textwidth}{@{}l >{\raggedright\arraybackslash}X >{\raggedright\arraybackslash}X@{}}
\toprule
Measure & Definition & What a positive value asserts \\
\midrule
SCR & Fraction accepted by \texttt{qasm3.loads} &
  The text parses \\
PCS & Declared parameters equal referenced parameters &
  The interface is closed \\
Full match & All structural fields agree (5 on QAOA, 6 on VQE) &
  The construction matches the reference \\
Skeleton match & Names quotiented; cover and slots only; QAOA &
  The name-free skeleton matches \\
Adherence & The artifact implements the contract it received &
  The plan was carried out \\
HQCR & $KL(P_{\mathrm{ref}}\|P_{\mathrm{arm}})\le 0.1$, floor-regularized &
  The output distribution is close \\
$q$-hit@$\tau$ & Fitted energy $\le E_{\mathrm{ref}}+\tau$ &
  The solution is within tolerance \\
\bottomrule
\end{tabularx}
\end{table*}

\subsection{Conditions}\label{sec:arms}

Seven conditions separate the contract from the checkpoint, and from the surface
form of the prompt; Table~\ref{tab:arms} lists them. Two comparisons are primary,
\ours{} against \oneshot{} and \ours{} against \ce{}. \ew{} shares the weights of
\ours{} so that a difference can be attributed to the contract, and it is no
performance baseline: with a placeholder in the contract slot it scores below the
one-stage condition. \goldc{} hands the generator the reference contract itself, which
the generator may still fail to execute. \renc{} and \shuf{} corrupt that contract; they are controls,
not methods.

An eighth condition, A1, controls the relevance of the contract rather than its
presence and runs after the seven: the conditioned generator receives a
shape-compatible but instance-irrelevant reference contract from another instance,
matched on family, qubit count and depth. Section~\ref{sec:relevance}
reports it with its result; it is not one of the seven conditions the main contrasts
use.

\begin{table*}[t]
\centering
\caption{Generation conditions: checkpoint, contract at inference and role. The primary
comparisons are Conditioned against One-stage and Conditioned against Matched-step.}
\label{tab:arms}
\scriptsize
\begin{tabularx}{\textwidth}{@{}l >{\raggedright\arraybackslash}X >{\raggedright\arraybackslash}X >{\raggedright\arraybackslash}X@{}}
\toprule
Condition & Checkpoint & Contract at inference & Role \\
\midrule
\oneshot{} & One-stage & None & Baseline \\
\ce{} & One-stage, 638 further steps & Placeholder & Matched step count \\
\ew{} & Shared with \ours{} & Placeholder & Equal weights \\
\ours{} & Mixture, below & Predicted from the instance & Contract \\
\goldc{} & Shared with \ours{} & Reference contract & Oracle reference \\
\renc{} & Shared with \ours{} & Reference, names bijection & Naming control \\
\shuf{} & Shared with \ours{} & Reference, grouping broken & Content control \\
\midrule
Predicted on plan-free weights & Without & Predicted contract & Crossed cell \\
Reference on plan-free weights & Without & Reference contract & Crossed cell \\
\midrule
A1 & Shared with \ours{} & Reference of another instance & Relevance control \\
Rule card & Shared with \ours{} & Deterministic rule on the term list & Planner-free card \\
\bottomrule
\end{tabularx}

\vspace{2pt}
{\footnotesize Equal-weight shares the Conditioned checkpoint. \goldc{} receives the
reference contract. Renamed and Shuffled corrupt it. The crossed cells are the same
prompts on weights continued without a contract (Table~\ref{tab:cell}). A1 and the
rule card are reported in Sections~\ref{sec:relevance} and~\ref{sec:attrib}.}
\end{table*}

Both generators are LoRA continuations of public Qwen2.5 checkpoints \citep{qwen,qwen25tr}. The
main comparison starts from Qwen2.5-3B-Instruct (snapshot \texttt{aa8e725}) and the scale
probe from Qwen2.5-7B-Instruct. Two stages precede the continuations: a base stage on
4{,}457 QAOA rows gives A0, and A0+ adds the variational family for 10{,}207 rows. A0+ is
the one-stage condition and the frozen ancestor of both continuations, and it never sees a
contract: its prompts carry no plan block at all. The planner is a separate LoRA on the
same 3B base, trained for two epochs on 9{,}604 instance-contract pairs at a learning rate
of $2\times10^{-4}$ and a maximum sequence length of 3{,}072. Training data for the
conditioned continuation mixes reference contracts, contracts produced by the planner, and
placeholder contracts, with measured shares 0.497, 0.302, and 0.201; the plan-free
continuation keeps the same block in every row and fills it with the placeholder. Both continuations that enter the primary contrast start from the same frozen
ancestor. They run 638 optimizer steps at an effective batch of 16 (one sequence per device
step, gradient accumulation 16), at learning rate $10^{-4}$ with a cosine schedule and
19 warmup steps, AdamW, maximum sequence length 4096, in bfloat16 on one RTX 4090~D
(24\,GB). The
adapters are LoRA of rank 16 on the seven attention and MLP projections.
Both corpora hold the same 9{,}604 rows, so the control fixes the schedule and the
instance count and varies only the corpus. The seed
is 42 for the frozen pair and 1337 for the replicate.

Planning is a separate forward pass over the instance text, into
whose prompt no energy, distribution, or reference parameter enters. Decoding is
greedy throughout. Renaming applies a group-preserving bijection to the symbol names
of the reference contract, drawn with a fixed seed; shuffling breaks the grouping of
terms across coefficient classes. The plan text of both controls is a deterministic
function of the reference contract and that seed, and, though not stored as a
separate log, the evaluation scripts regenerate it.

\subsection{Statistical analysis}\label{sec:stats}

Comparisons are paired: an exact McNemar test on the discordant pairs, and a 95\%
percentile paired bootstrap interval at 4{,}000 resamples, which no bound moves by
more than 0.3 points when recomputed at 20{,}000. Every bootstrap interval in the
paper, including those at 20{,}000 resamples, is of that form. Every cell reports its $n$. Each test compares paired outcomes on this fixed set. It is not an inference to a superpopulation, and the set is the constructed sample of Section~\ref{sec:instances}. The exact McNemar test is the primary
inference and the bootstrap interval is descriptive: where the two disagree, as on the
narrow conjunction of cover and slot against the reference contract, the exact test
governs and the bootstrap interval is reported beside it.

Three analyses are primary and the performance claim rests on those alone, all on QAOA
full match: Conditioned against One-stage, Conditioned against Matched-step, and the
interaction of Table~\ref{tab:cell}, which is the estimate the two-by-two was built for.
A Holm correction over the three leaves all three significant (adjusted
$p \le 7\times10^{-5}$). The other rows of
Table~\ref{tab:paired} compare attribution and control conditions; they carry exact
tests and intervals, they are left uncorrected, and none is read on its own. The same convention applies to the semantic contrasts,
which carry their discordant counts in Section~\ref{sec:semantics}, and to the
overlap views, which are the same contrast measured three ways rather than three
independent tests.
Each primary contrast is also reported on three overlapping views of
training--evaluation overlap, and the strictest view is treated as underpowered
rather than as a negative result (Section~\ref{sec:contam}). Semantic comparisons
use the rows on which the shared refit converged, so their $n$ is stated with the
contrast.

\section{The construction contract}\label{sec:criterion}

\subsection{What the contract is for}\label{sec:reduction}

A construction contract is an intermediate, machine-checkable specification of the
structural decisions a generated artifact is expected to realize. The case for putting it
before generation is timing and price. A score of the finished artifact is computed after
the last token, it is one number for the whole program, and on this benchmark it costs
about 36\,s per QAOA artifact and 3.4--4.2\,min per VQE artifact once the variational refit
is included. A dropped term stays invisible at the step where it was dropped. The contract
moves a checkable description of the construction to the input side, where it costs one
forward pass.

There is no precondition, no postcondition, and no runtime check
(Section~\ref{sec:related}). The object is a construction summary, written before
the program exists and compared with the program afterward. Formally, a contract is a
value in a product $S=S_1\times\cdots\times S_k$ in which each factor is constant or linear in
the number of qubits. An extractor
$\varphi$ reads an artifact and returns the contract it realizes. A planner $\psi$ reads the instance and returns a contract. The extractor is a
deterministic walk over the text; the planner is a learned forward pass. The
extractor does not call a simulator. Full match and adherence are then
indicators of equality in $S$,
\begin{equation}
\label{eq:full}
\mathrm{full\ match}(p)=\mathbf{1}[\varphi(p)=\varphi(p^{\star})],
\end{equation}
\begin{equation}
\label{eq:adh}
\mathrm{adherence}(p,s)=\mathbf{1}[\varphi(p)=s],
\end{equation}
where $p^{\star}$ is the reference artifact and $s$ is the contract the
generator was given. Two of the five QAOA fields are the contract's own. Order
coincides with cover on this benchmark, and inversion is carried as a check.
Rename and shuffle are what separate full match from skeleton match
(Table~\ref{tab:fields}). Equation~\eqref{eq:adh} would accept a generator that
implements a wrong contract, which is why Equation~\eqref{eq:full} is primary.

Skeleton match is the same indicator after quotienting the names. Let $\rho$ be the
relabelling that maps each symbol to its rank in the order of first appearance in
the artifact, and let $\pi$ keep only the cover and the sequence of slots, so that
$\sigma=\pi\circ\rho$ sends a contract to its name-free skeleton. Then
\begin{equation}
\label{eq:skel}
\mathrm{skeleton\ match}(p)=\mathbf{1}[\sigma(\varphi(p))=\sigma(\varphi(p^{\star}))],
\end{equation}
defined on QAOA only, because a variational contract carries no symbol that a
relabelling could act on. Equation~\eqref{eq:skel} is invariant under a
group-preserving renaming of the reference by construction. That invariance is what makes the Renamed condition of Table~\ref{tab:arms} a control on the naming rather than on the construction.

\subsection{Reading the two families after the fact}\label{sec:properties}

Three descriptions are reported, because the measurements below do not mean what they
say without them. They were formulated after looking at this benchmark and are
evaluated on it retrospectively; none of them predicts which family pays. Each is
measured on both families in Table~\ref{tab:criteria}.

Two are properties of the object. \emph{Decidability}: every field is settled by a walk
over the text, with no semantic oracle, and 99.97\% of the reference corpus falls under
the extraction schema. \emph{Diagnosability}: failures land in named fields, so a miss
can be attributed; the coefficient slot is missed in 78\% of the failed QAOA artifacts
and the covered terms in 38\%, and on VQE the repetition count in 93\% and the motif in
73\%. The third is a property of the pipeline. \emph{Realizability}: the generator can
carry a reference contract out at all. Adherence is 0.808 on QAOA and 0.020 on VQE at
3B, and the variational family is the case in which a realizable contract is not
available (Section~\ref{sec:vqe}).

Planner accuracy was considered and is not a requirement. The planner leads one-stage
on the scored fields: 0.930 and 0.850 against 0.780 and 0.570 on QAOA, and 0.630 and
0.480 against 0.070 and 0.200 on VQE. A deterministic rule supplies a card of the same
value, so the effect is not planner accuracy (Section~\ref{sec:attrib}). It is
card-conditioned training, which that section measures and does not decompose. The
three readings are a retrospective description of these data, not a forecast
(Section~\ref{sec:th-conclusion}).

\begin{table*}[t]
\centering
\caption{Three retrospective readings (Section~\ref{sec:properties}): what each one judges, how
it is measured, and what the two families show.}
\label{tab:criteria}
\small
\setlength{\tabcolsep}{4pt}
\begin{tabularx}{\textwidth}{@{}l l >{\raggedright\arraybackslash}X >{\centering\arraybackslash}p{2.1cm} >{\centering\arraybackslash}p{3.1cm}@{}}
\toprule
Reading & Object & How it is measured & QAOA & VQE \\
\midrule
Decidability & the contract & Coverage of the extraction schema
  & \multicolumn{2}{c}{99.97\% of the corpus} \\
Diagnosability & the error distribution
  & Share of failed artifacts in which a decision field is missed
  & \begin{tabular}[c]{@{}c@{}}slot 78\%\\cover 38\%\end{tabular}
  & \begin{tabular}[c]{@{}c@{}}repetition 93\%\\motif 73\%\end{tabular} \\
Realizability & the generator
  & Adherence to the card received: predicted card, then reference card
  & \begin{tabular}[c]{@{}c@{}}0.808\\0.870\end{tabular}
  & \begin{tabular}[c]{@{}c@{}}0.020\\0.030\\utility not estimable\end{tabular} \\
\bottomrule
\end{tabularx}

\vspace{2pt}
{\footnotesize Coverage uses 10{,}207 reference circuits. Adherence compares decisions
the card carries. Allocation, order, and inversion are checked only by full match.
Near-zero adherence leaves the utility of realized plans not estimable.}
\end{table*}

Table~\ref{tab:criteria} is the reading used below. On the one-stage arm alone the
QAOA slot is still missed in 87.8\% of failures and the covered terms in 44.9\%, so
the concentration is not an artifact of arms built to damage a field. Handing VQE the
reference card yields full match 0.030 while cutting parse from 0.810 to 0.550. A
length-matched random card lands near that interface loss (parse 0.520), so much of
the break tracks the volume of the block. Semantic contrasts on VQE would score a
contract the model rarely executes.

\subsection{The contract used in the study}\label{sec:contract}

For QAOA each line of the contract names an operation, the set of terms it
covers, and the coefficient slot of its symbol. Line order is recorded with
the line. For VQE a line names the motif, the repetition count, the sharing
pattern, and the tail, while the gate set and the arity are recorded in the header
and checked once per artifact; all six are part of the predicate. The same header and the same line index are used for
both families, so a comparison can be made line by line even though the fields
differ. Fig.~\ref{fig:example} shows the relation on one instance: a parsed
artifact can still miss terms, mis-assign depth, or fail to share symbols, and
the contract is the short object that records those choices.

\begin{figure*}[t]
\centering
\begin{minipage}[t]{0.31\linewidth}
{\ttfamily\footnotesize
\textrm{\textbf{reference construction}}\\[3pt]
CARD qaoa n=5 p=1\\
1 cost [0,1,2] x0\\
2 cost [3,4,5,6,7,8] x1\\
3 cost [9] x2\\
4 lin [10,11,12] x3\\
5 lin [13,14] x4\\
6 \textcolor{red}{mixer [0,1,2,3,4] x5}\par}
\end{minipage}\hfill
\begin{minipage}[t]{0.31\linewidth}
{\ttfamily\footnotesize
\textrm{\textbf{conditioned artifact}}\\[3pt]
OPENQASM 3.0;\\
include "stdgates.inc";\\
input float[64] x4;\\
\textcolor{red}{input float[64] x5;}\\
qubit[5] q;\\
\textcolor{black!45}{$\cdots$}\\
h q[0]; \textcolor{red}{rz(x5)} q[0]; h q[0];\\
h q[1]; \textcolor{red}{rz(x5)} q[1]; h q[1];\\
\textcolor{black!45}{$\cdots$}\par}
\end{minipage}\hfill
\begin{minipage}[t]{0.31\linewidth}
{\ttfamily\footnotesize
\textrm{\textbf{one-stage artifact}}\\[3pt]
OPENQASM 3.0;\\
include "stdgates.inc";\\
input float[64] x4;\\
\textcolor{red}{\textit{(no x5 declared)}}\\
qubit[5] q;\\
\textcolor{black!45}{$\cdots$}\\
h q[0]; \textcolor{red}{rz(x4)} q[0]; h q[0];\\
h q[1]; \textcolor{red}{rz(x4)} q[1]; h q[1];\\
\textcolor{black!45}{$\cdots$}\par}
\end{minipage}
\caption{One instance, three objects, reproduced from the per-row logs. Left: the
contract extracted from the reference. Center and right: excerpts from the Conditioned
and One-stage artifacts. Both parse and close the interface. Red marks the
coefficient-slot difference; gray marks omitted lines.}
\label{fig:example}

\end{figure*}

\begin{figure}[t]
\centering
\begin{tikzpicture}[
  node distance=5mm,
  box/.style={draw, rounded corners=1pt, align=center, inner sep=3pt,
              font=\footnotesize, minimum height=7mm},
  lab/.style={font=\scriptsize\itshape, fill=white, inner sep=1pt},
  arr/.style={->, >=stealth, semithick}]
\node[box, text width=4.6cm] (inst) {instance text\\[-1pt]
  {\scriptsize $n$, $p$, terms, coefficients}};
\node[box, below=of inst, text width=3.4cm] (card) {construction contract\\
  {\scriptsize 76--1{,}489 characters}};
\node[box, below=of card, text width=5.2cm] (art) {artifact: OpenQASM 3\\
  {\scriptsize 438--9{,}272 characters}};
\node[box, below=of art, text width=5.2cm, fill=black!5] (chk)
  {structural checker, no simulator\\[-1pt]
   {\scriptsize cover, slot, order, inversion}};
\draw[arr] (inst) -- node[lab, right] {planner, 1 pass} (card);
\draw[arr] (card) -- node[lab, right] {generator, 1 pass} (art);
\draw[arr] (art) -- node[lab, right] {no fit, no run} (chk);
\node[box, dashed, right=5mm of art, text width=2.0cm, draw=black!45, text=black!50,
      align=center] (unused) {\scriptsize\itshape out of path:\\[-1pt]
      simulator / solver\\[-1pt] / retrieval};
\end{tikzpicture}
\caption{Inference path of contract-conditioned generation. One forward pass predicts the
contract from the instance text, a second generates the artifact from instance and
contract, and the checker compares contracts by a text walk. The dashed box is deliberate:
no simulator, classical solver, or retrieval step is used on this path
(Section~\ref{sec:measures}).}
\label{fig:pipeline}
\end{figure}
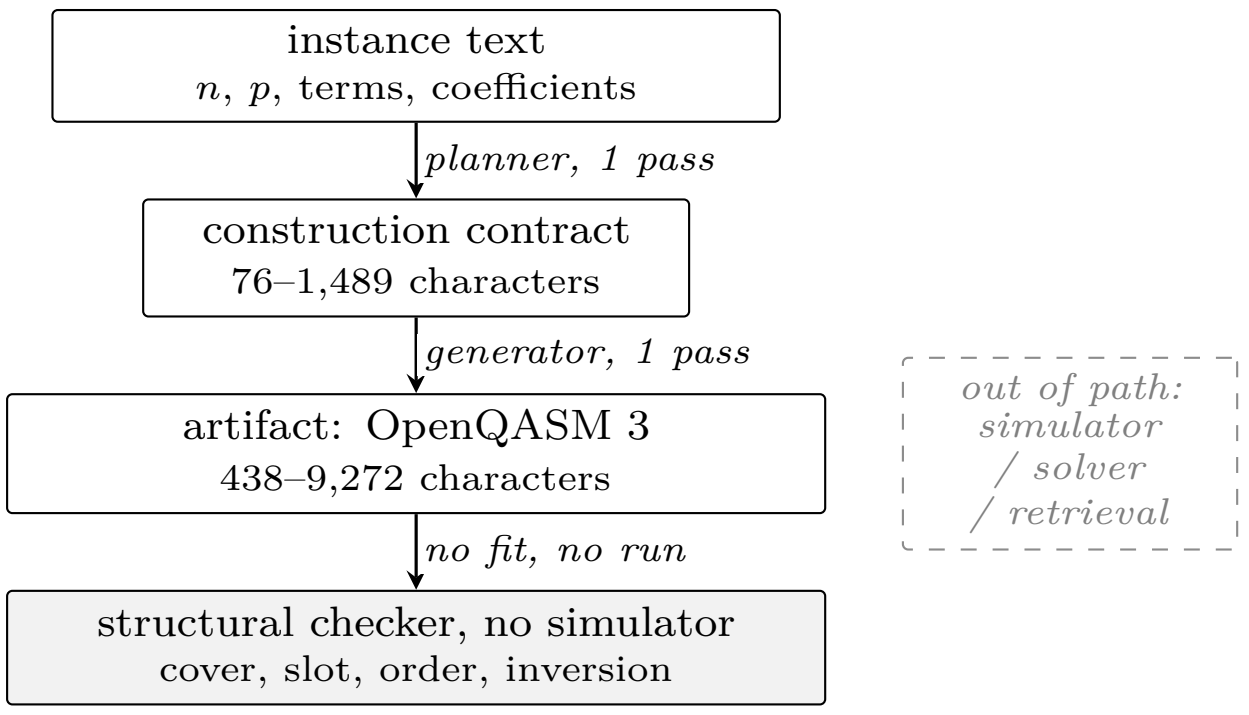

A predicted contract is 76--1{,}489 characters, with a median of 336 on
QAOA and 101 on VQE. The artifact it describes is 438--9{,}272 characters, with a
median of 1{,}502 and 3{,}707. Counts are of the emitted text, so they can be
checked against the per-row logs. Fig.~\ref{fig:pipeline} shows the inference path. The
planner reads the instance. The generator reads the instance and the contract.
The checker compares contracts and does not call a simulator.

\subsection{How the fields were chosen}\label{sec:fields}

The field list was fixed from the reference corpus before any condition was
evaluated, by three filters that use the reference artifacts alone. Coverage removes the
three circuits the schema cannot represent. Invariance drops a field that no reference
circuit violates: allocation and inversion leave the decision list at this step, and
ordering leaves separately, because the order of a reference circuit is always the order
of its cover. Referential closure keeps the fields a walk over the text can settle, which
is what excludes anything the variational evaluator would have to check. The test is a
count over all 10{,}207 circuits rather than a threshold on a sample, so the resulting
list does not depend on how the corpus is split. All three dropped fields remain in the
predicate as checks: the checker certifies them, and Section~\ref{sec:fields-fail} reports
how often the generator misses them anyway.

Table~\ref{tab:fields-select} records the role of each candidate. A decision is covered,
unviolated by the references, and settled by a walk. A check is unviolated by the
references. A field is excluded when the variational evaluator would have to verify it.
The absent parameter budget is taken up in Section~\ref{sec:vqe}.

\begin{table*}[t]
\centering
\caption{Why each candidate field is a decision, a check, or absent. The evidence column
records what each field was judged on before any condition was evaluated.}
\label{tab:fields-select}
\footnotesize
\begin{tabular}{@{}p{2.4cm}p{1.6cm}p{7.2cm}p{2.4cm}@{}}
\toprule
Field & Role & Evidence & Decision \\
\midrule
cover (QAOA) & decision & covered by the schema; decidable by a walk; varies in the reference corpus & kept \\
slot (QAOA) & decision & covered by the schema; decidable by a walk; varies in the reference corpus & kept \\
motif (VQE) & decision & covered by the schema; decidable by a walk; varies in the reference corpus & kept \\
repeat (VQE) & decision & covered by the schema; decidable by a walk; varies in the reference corpus & kept \\
allocation & check & never violated by a reference circuit & kept as a check \\
ordering & check & never violated by a reference circuit; equal to cover in every condition & kept as a check \\
inversion & check & never violated by a reference circuit & kept as a check \\
\texttt{rx} gate & excluded & never emitted (0\% of artifacts) & removed \\
parameter budget & absent & arms emit median 93 parameters against 42 & not carried; failure in Section~\ref{sec:vqe} \\
\bottomrule
\end{tabular}

\vspace{2pt}
{\footnotesize Filters use the reference corpus only, before any condition is evaluated.
Failure location (Section~\ref{sec:properties}) did not choose the list. Order equals
cover, so the five QAOA fields are four verdicts plus one check.}
\end{table*}

What survives is a symbol convention read off the corpus rather than chosen: the median
number of declarations is 7 against 13 terms, symbols are shared by coefficient class, and
\texttt{rx} is never emitted. Two later observations are reported only as description.
Failures concentrate in the coefficient slot and the covered terms
(Section~\ref{sec:properties}), and a placeholder block on the same checkpoint reaches
full match 0.460 against 0.510 for one-stage. Neither observation entered the selection:
the list was fixed first, and the concentration was measured afterward.

\subsection{Alternatives we rejected}\label{sec:alternatives}

Before settling on the contract we considered three other reductions of the task. A vector
of fitted parameters fails decidability: comparing two vectors reintroduces the variational
fit the reduction was meant to avoid. Per-block shares fail for the same reason. Solution
support fails diagnosability, because two artifacts may cover the same terms and still
differ in depth, order, and symbol sharing, so a support mismatch names no failing step.
Only the contract met all three readings of Section~\ref{sec:properties}, and it pays for
them with a family-specific field list (Section~\ref{sec:disc-transfer}).

\section{Results}\label{sec:results}

All conditions use the same 200 instances, greedy decoding, and no post-hoc repair.
They differ only in the checkpoint and in the contract-slot text
(Section~\ref{sec:arms}). Each cell is a paired McNemar test with a bootstrap interval
(Section~\ref{sec:stats}). Structural rates are read from the emitted text. Semantic
measures use the shared refit and are interpreted on QAOA. The subsections follow
Section~\ref{sec:rq}.

\subsection{Parsing and structural match}\label{sec:parse}

Question~1 asks where the artifacts fail, and whether a semantic score taken on the
emitted parameters can see it. Table~\ref{tab:main} reports the seven conditions on QAOA:
parse lies between 0.790 and 0.970, full match between 0.400 and 0.870. Across both
families the ranges widen to $[0.55,0.97]$ and $[0.02,0.87]$. The extreme cell is on VQE
(Section~\ref{sec:vqe}): one-stage parses 0.810 of those artifacts and matches all six
fields on 0.020.

\begin{table*}[t]
\centering
\caption{Structural and semantic results on QAOA over the seven conditions, 100 instances
per condition. Full match is the primary structural measure.}
\label{tab:main}
\footnotesize
\begin{tabular}{@{}lrrrrrr@{}}
\toprule
Condition & SCR & PCS & Full & Skel. & Adhere & HQCR \\
\midrule
\oneshot{} & 0.940 & 0.790 & 0.510 & 0.570 & --- & 0.670 \\
\ce{}      & 0.910 & 0.770 & 0.550 & 0.630 & --- & 0.780 \\
\ew{}      & 0.870 & 0.720 & 0.460 & 0.550 & --- & 0.724 \\
\shuf{}    & 0.790 & 0.670 & 0.400 & 0.660 & --- & 0.683 \\
\renc{}    & 0.930 & 0.680 & 0.530 & 0.710 & --- & 0.753 \\
\ours{}    & 0.970 & 0.900 & 0.780 & 0.840 & 0.808 & 0.856 \\
\goldc{}   & 0.970 & 0.940 & 0.870 & 0.880 & --- & 0.938 \\
\bottomrule
\end{tabular}

\vspace{2pt}
{\footnotesize SCR is the parse rate and PCS the parameter-interface rate. Skeleton
match is defined on QAOA only. Adherence is computed for Conditioned ($n=99$: one
artifact could not be extracted). Cells marked --- are undefined. HQCR uses converged
rows only, $n=94,91,87,79,93,97,97$ in row order. Order equals cover on this family
(Table~\ref{tab:fields}).}
\end{table*}

A score taken on the parameters as emitted carries none of that spread. With every symbol
bound to the constant 0.1, and the reference circuit bound the same way, the median
absolute energy error and the median distributional divergence are zero for every
condition, including those at full match 0.400 and 0.870. The protocol is degenerate by
construction; the zero medians are reported once, here, as the reason the semantic
measures are computed after the shared refit rather than read off the emitted text
(Section~\ref{sec:semantics}).

\emph{Answer to Question~1.} On QAOA the failures are structural rather than syntactic.
On VQE the carded cells lose parse as well (0.550 against 0.810), while full match stays
between 0.02 and 0.04 either way. The emitted-parameter score ranks nothing.

\subsection{Effect of the contract}\label{sec:effect}

Question~2 is answered on QAOA by two paired comparisons of full match, both with
$n=100$, on the constructed set of Section~\ref{sec:instances}. Eighty of 100 QAOA
signatures appear in some training split. The 20 rows absent from every split are in
Table~\ref{tab:strata}, and that interval includes zero.

Conditioned reaches 0.780 and one-stage 0.510: $+27.0\pt$, interval $[+17,+37]$,
McNemar $p=1.4\times10^{-6}$. That contrast changes the checkpoint as well as the
prompt, so $+27.0$ is the pipeline a practitioner deploys. Matched-step continues the
same ancestor for the same 638 steps on the plan-free corpus and receives a
placeholder. It reaches 0.550. Against Conditioned the difference is $+23.0\pt$,
interval $[+13,+33]$, $p=3.4\times10^{-5}$. Ancestor, step count, and instance count
are held fixed. The corpus is what differs. Both gains exceed the spread of the parse
rate on this family, so neither is a parsing effect. The estimate is conditional on
the set as constructed
(Sections~\ref{sec:contam} and~\ref{sec:th-conclusion}).

\begin{figure}[t]
\centering
\includegraphics[width=\linewidth]{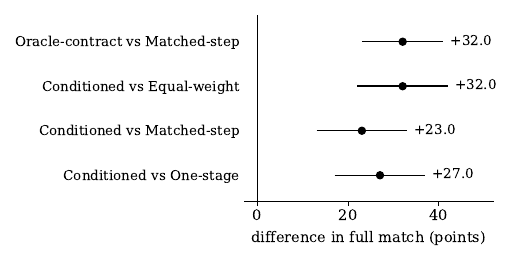}
\caption{Performance and attribution contrasts, plotted from the numbers of
Table~\ref{tab:paired}. Points are paired differences on QAOA full match, bars are paired
bootstrap intervals, and the vertical line is no difference.}
\label{fig:contrasts}

\end{figure}

Equal-weight holds the conditioned weights and empties the contract slot. Full match
moves by $+32.0\pt$ and skeleton match by $+29.0\pt$. The checkpoint cannot account
for either difference, and $+32.0\pt$ is larger than either trained contrast.
Fig.~\ref{fig:contrasts} plots the intervals. The control
holds the schedule fixed rather than the prompt-token count, because a contract
occupies more of the prompt than a placeholder. The whole plan-free continuation is
worth $+4.0\pt$ against the uncontinued base, interval $[-3,+11]$, $p=0.42$.

The 100 QAOA rows carry 98 distinct signatures. A cluster bootstrap (20{,}000
resamples) and a sign-flip test on cluster means leave the primary contrast at
$+23.0\pt$, interval $[+13.3,+33.7]$, $p<0.001$. No other primary bound moves by
more than 0.7 points.

Recomputed on covered terms and coefficient slots, the two fields the contract
decides, Conditioned reaches 0.840 against 0.570 for one-stage ($+27.0\pt$,
$[+18,+36]$), $+21.0\pt$ against matched-step ($[+13,+30]$), and $+29.0\pt$ at fixed
weights ($[+19,+39]$). Against matched-step the full-match difference is
$+23.0\pt$, so the fields outside the contract account for those two points. The gap between the learned planner and the reference
contract falls from $9.0\pt$ to $4.0\pt$ on that conjunction, and four discordant
pairs do not establish it (Section~\ref{sec:attrib}). Of 22 conditioned mismatches,
16 fail on a covered term or a slot and 6 only on assignment or order. Matched-step
splits 37 and 8 of 45 the same way.

Re-training both continuations from the same ancestor with seed 1337 reaches 0.770
against 0.570 ($+20.0\pt$, $[+10.0,+30.0]$, $p=3.2\times10^{-4}$). Seed 2026
reaches 0.770 against 0.590 ($+18.0\pt$, $[+9,+28]$, $p=9.1\times10^{-4}$). The
three seeds give $+23.0$, $+20.0$ and $+18.0$. The base checkpoint and the planner
stay fixed, so the replicate covers the continuation only.

A length-matched random card keeps the predicted card's length and character classes
and replaces every alphanumeric character with a seeded random string. It parses
0.730, closes the interface on 0.430, and reaches full match 0.270 against 0.780:
$+51.0\pt$, 52 discordant rows against 1, interval $[+41,+61]$,
$p=1.2\times10^{-14}$. It is longer than the placeholder on the same weights
(0.460), so the length comparison is conservative. On VQE the same control reaches
0.020 against 0.030.

The gain is not uniform. Splitting the 100 QAOA pairs by the length of the
conditioned program, a split on a treatment outcome and therefore a mechanism
check, gives $+5.6\pt$, $+41.9\pt$ and $+24.2\pt$ in the shortest, middle, and
longest thirds against Matched-step. Matched artifacts are shorter (median 104
lines against 146).

A fixed-weight probe continues Qwen2.5-7B-Instruct on the same 9{,}604-row
plan-conditioned corpus for 638 steps from a fresh LoRA. It has no base stage, no
one-stage arm, and no plan-free continuation, so it is a second scale, not a
replication. On QAOA, reference card, predicted card, and placeholder give full match
0.920, 0.770 and 0.250. Adherence is 0.930 and 0.808, the latter equal to the 3B arm.
On the 42 QAOA rows where both conditions close the interface, the predicted card
reaches 0.952 against 0.548 for the placeholder. That check conditions on an outcome.
The variational half does not move (Section~\ref{sec:vqe}). Pooled over 200 instances,
Conditioned against Matched-step is $+12.0\pt$ $[+7.0,+17.5]$.
Table~\ref{tab:paired} collects the paired tests.

\emph{Answer to Question~2.} The same-step gain is $+23.0\pt$ and the pipeline gain
is $+27.0\pt$. Only the equal-weight contrast ($+32.0\pt$) holds the checkpoint
fixed. The estimate is conditional on the constructed QAOA set.

\begin{table*}[t]
\centering
\caption{Paired comparisons on QAOA. Unless a row says otherwise, $n=100$. Full match is
the primary measure. Disc.\ counts the discordant pairs of the exact test.}
\label{tab:paired}
\scriptsize
\setlength{\tabcolsep}{4pt}
\begin{tabular}{@{}lrlrl@{}}
\toprule
Comparison & Full match [interval] & $p$ & Skeleton [interval], $p$ & Disc. \\
\midrule
\ours{} vs \oneshot{} & $+27.0$ $[+17,+37]$ & $1.4\times10^{-6}$ & $+27.0$ $[+18,+37]$, $1.1\times10^{-7}$ & 30/3 \\
\ours{} vs \ce{} & $+23.0$ $[+13,+33]$ & $3.4\times10^{-5}$ & $+21.0$ $[+13,+29]$, $5.7\times10^{-6}$ & 27/4 \\
\ours{} vs \ew{} & $+32.0$ $[+22,+42]$ & $6.7\times10^{-8}$ & $+29.0$ $[+20,+39]$, $1.3\times10^{-7}$ & 35/3 \\
\goldc{} vs \ce{} & $+32.0$ $[+23,+41]$ & $4.7\times10^{-10}$ & $+25.0$ $[+17,+34]$, $6\times10^{-8}$ & 32/0 \\
\ours{} vs \goldc{} & $-9.0$ $[-15,-4]$ & $0.0039$ & $-4.0$ $[-8,-1]$, $0.12$ & 0/9 \\
\renc{} vs \goldc{} & $-34.0$ $[-44,-24]$ & $5.4\times10^{-9}$ & $-17.0$ $[-25,-9]$, $7.6\times10^{-5}$ & 2/36 \\
\shuf{} vs \ew{} & $-6.0$ $[-19,+8]$ & $0.47$ & $+11.0$ $[-1,+23]$, $0.11$ & 21/27 \\
Rule card vs \ours{} & $-6.0$ $[-14,+2]$ & $0.24$ & --- & 6/12 \\
\midrule
Executor vs \ours{} (predicted) & $-2.0$ $[-7,+3]$ & $0.69$ & --- & 2/4 \\
Executor vs rule arm & $+6.0$ $[+1,+12]$ & $0.070$ & --- & 7/1 \\
\midrule
Replicate (seed 1337) & $+20.0$ $[+10,+30]$ & $3.2\times10^{-4}$ & --- & 25/5 \\
Replicate (seed 2026) & $+18.0$ $[+9,+28]$ & $9.1\times10^{-4}$ & --- & 23/5 \\
\bottomrule
\end{tabular}

\vspace{2pt}
{\footnotesize The two $-6.0$ rows are different contrasts: Shuffled versus
Equal-weight is a content control on a corrupted reference card; Rule card versus
\ours{} is the planner-free card against the learned planner's card on the same
conditioned weights (Section~\ref{sec:attrib}). Skeleton match is not scored for
the rule or executor rows.}
\end{table*}

\begin{table}[t]
\centering
\caption{Two-by-two on QAOA full match, $n=100$. Rows are the second-stage training.
Columns are the contract slot at inference.}
\label{tab:cell}
\footnotesize
\setlength{\tabcolsep}{3pt}
\begin{tabular}{@{}lccc@{}}
\toprule
 & placeholder & predicted & reference \\
\midrule
with a card & 0.460 & 0.780 & 0.870 \\
without a card & 0.550 & 0.420 & 0.450 \\
\bottomrule
\end{tabular}

\vspace{2pt}
{\footnotesize The lower-right cells hand a plan to weights that never saw a contract.}
\end{table}

\subsection{What the gain depends on}\label{sec:attrib}

Question~3 asks which part of the gain survives once the checkpoint, the step budget,
the formatted block, and the card content are separated.

The shared checkpoint is not sufficient. Emptying the slot on the conditioned weights
drops full match from 0.780 to 0.460 ($+32.0\pt$, $[+22,+42]$, $p=6.7\times10^{-8}$).
The re-trained pair gives $+31.0\pt$ ($[+20,+42]$, $p=3.4\times10^{-7}$). The
placeholder lies below one-stage (0.460 against 0.510) and is not a performance
baseline. Extra steps are not sufficient either: Matched-step reaches 0.550.

Table~\ref{tab:cell} is the two-by-two the design was built for. On weights trained
with a card the contract is worth $+32.0\pt$ (35/3, $[+22,+42]$, $p=6.7\times10^{-8}$).
On weights trained without one the same card costs $13.0\pt$ against their own
placeholder (8/21, $[-23,-3]$, $p=0.024$), and the reference contract costs $10.0\pt$
(8/18, $[-20,+0]$, $p=0.076$). The interaction, the difference of those simple
effects, is $+45.0\pt$, interval $[+36,+62]$, sign-flip $p<5\times10^{-5}$. The
reference-contract cell points the same way, but its interval ends at zero, so the
predicted-contract cell carries that half of the inference.

The contrast establishes that the effect requires training on card-conditioned data.
It does not separate which field the generator exploits. The conditioned corpus mixes
reference, predicted, and placeholder cards in the shares 0.497, 0.302, and 0.201.
The plan-free corpus contains none. Every row of both corpora carries the plan block,
so the format is held fixed. A placeholder on conditioned weights (0.460) and a
contract on plan-free weights (0.420--0.450) both fall below the one-stage baseline of
0.510. Neither training nor a formatted block substitutes for the pair.

Names are load-bearing. A group-preserving renaming of the reference contract costs
$17.0\pt$ on skeleton match ($p=7.6\times10^{-5}$) and $34.0\pt$ on full match
($p=5.4\times10^{-9}$), both against \goldc{}. Breaking the cross-class grouping
yields full match 0.400, below the placeholder at 0.460, but the paired comparison is
$-6.0\pt$ ($[-19,+8]$, $p=0.47$). The skeleton comparison on the same rows is
$+11.0\pt$ ($p=0.11$). The sharper content evidence is the relevance control: a card
of another instance is worth 0.130 against the placeholder's 0.460
(Section~\ref{sec:relevance}).

A deterministic rule can write a card of comparable value. It groups consecutive
terms of equal coefficient into one block and names each block by its coefficient
class. It reproduces the reference cover in 94 of 100 QAOA rows and the whole
decision vector in 73. On the 34 rows where the rule and the reference disagree, the
generator follows the card in 4, the reference in 17, and neither in 13. Handed to
the conditioned generator, the rule card reaches full match 0.720 and is implemented
in 0.700 of rows. The predicted contract reaches 0.780 (difference $-6.0$,
$[-14,+2]$, $p=0.24$, discordant $6/12$). The placeholder reaches 0.460 ($+26.0$,
$[+16,+36]$, $p=6.2\times10^{-6}$, $30/4$) and the reference contract 0.870
($-15.0$, $[-22,-8]$, $p=6.1\times10^{-5}$). The weights still have to know how to
read a card. The rule was not scored on plan-free weights.

\emph{Answer to Question~3.} The contract has to be in the prompt, and the generator
has to have been trained with one. Either alone leaves full match at or below the
one-stage baseline. A legal but unrelated plan is worse than none. On these 100 rows
a rule that reads the term list is not distinguished from the planner.

The executor reads the card and emits the fixed gate template of this corpus, one
block per step, with no model in the loop. Fed the predicted card it reaches full
match 0.760 and reproduces the card on 89 of the 92 it can emit. The generator
reaches 0.780, a difference of $-2.0$ (2/4 discordant, $[-7.0,+3.0]$, $p=0.69$).
Non-inferiority is not established: the interval admits a disadvantage of about 7
points. Fed the rule card the executor reaches 0.780 against 0.720 ($+6.0$, 7/1,
$[+1.0,+12.0]$, $p=0.070$). The eight predicted cards that cannot be emitted are
counted as executor failures. The executor is a QAOA baseline. A variational
contract does not determine a gate sequence in the same way, and none is run there.

\goldc{} remains above Conditioned. Full match falls by $9.0\pt$ from 0.870 to
0.780 ($[-15,-4]$, $p=0.0039$). On the narrower conjunction the same comparison is
$-4.0\pt$ from four discordant pairs. The exact test gives $p=0.125$, while the
bootstrap interval $[-8,-1]$ excludes zero. The residual is visible on the stricter
predicate and is not established on the narrower one.

\subsection{The family on which the defined card fails}\label{sec:vqe}

Question~4 compares the two families on realizability, the reading that separates them in
Table~\ref{tab:criteria}. On QAOA the conditioned generator carries the reference
contract out (adherence 0.808) and full match rises by 27.0 points. On VQE adherence is
0.020, so the generator does not carry this card out.
Table~\ref{tab:vqe} shows the consequence. Full match sits between 0.020 and 0.040 for every 3B condition. \goldc{}, which is given the reference contract,
reaches full match 0.030 and parses only 0.550
of its outputs, the lowest parse rate in the table. The card as defined here is rarely
realized (adherence 0.020 under Conditioned). Supplying the reference card does not raise
full match, so planner field accuracy alone is not a sufficient repair; whether a different
field list---in particular one that carries a parameter budget---would change the outcome
is untested and not separated by this design.

\begin{table*}[t]
\centering
\caption{The VQE family over the same seven conditions, 100 instances per condition. Skeleton
match is undefined here and is omitted.}
\label{tab:vqe}
\footnotesize
\begin{tabular}{@{}lrrrrrrr@{}}
\toprule
Condition & SCR & PCS & SCR$\wedge$PCS & Full match & Adherence & $q$-hit@0.2 & HQCR \\
\midrule
\oneshot{} & 0.810 & 0.810 & 0.770 & 0.020 & ---   & 0.296 & 0.037 \\
\ce{}      & 0.700 & 0.670 & 0.650 & 0.020 & ---   & 0.286 & 0.043 \\
\ew{}      & 0.790 & 0.720 & 0.720 & 0.030 & ---   & 0.184 & 0.026 \\
\shuf{}    & 0.650 & 0.590 & 0.590 & 0.040 & ---   & 0.185 & 0.031 \\
\renc{}    & 0.810 & 0.730 & 0.730 & 0.030 & ---   & 0.198 & 0.037 \\
\ours{}    & 0.690 & 0.640 & 0.630 & 0.030 & 0.020 & 0.290 & 0.029 \\
\goldc{}   & 0.550 & 0.470 & 0.470 & 0.030 & ---   & 0.182 & 0.054 \\
\bottomrule
\end{tabular}

\vspace{2pt}
{\footnotesize PCS is the parameter-interface rate. SCR$\wedge$PCS is the joint event.
Adherence is computed for Conditioned. The reference card leaves full match at 0.030,
so the miss is not only planner error.}
\end{table*}

The two semantic columns on this family are descriptive of the emitted artifacts,
not a semantic return of the contract: they are computed from the artifacts that
were produced and refitted under the shared budget, and with adherence at 0.020 no
contract-side benefit is defined here. Only one energy comparison remains estimable
on this family: whether VQE became
worse. On the paired subset of $n=45$, \goldc{} against Matched-step on $q$-hit@0.2 is
$-8.9\pt$, with interval $[-22.2,+4.4]$ and $p=0.34$. The interval excludes a
gain of about 15 points and includes a degradation of about 22 points. The
study draws no conclusion of benefit or of harm on VQE energy. What it does
locate is the failed step. Two measurements are consistent with an execution
failure. Conditioned and \goldc{} emit a median of 93 parameters against 42 in the
reference circuits, and the contract carries no field for that budget. The
corpus spans 17 ansatz families, of which the evaluation set represents two,
and the motif field on the \goldc{} condition is only 0.410.

Whether that failure remains at the 7B scale is left open. On the 7B probe the
reference contract lifts variational full match from 0.050 to 0.210 on 16 discordant
pairs, while the predicted contract does not lift it (0.070 against 0.050, 5/3); the four
variational cells are in Table~\ref{tab:7b-vqe}. Two relations there are worth stating.
First, on the reference card adherence and full match coincide in all 100 rows, because
the card and the reference construction are one object. On the QAOA reference cell they
coincide in 99 of 100 rows; the one row that differs reproduces the card's construction
\emph{and} allocates its lines correctly, but references a parameter it never declares,
so it breaks two interface invariants that full match checks and the card does not carry.
A decision-level adherence measure is therefore not a weaker copy of full match: it is the
same predicate minus the fields the contract does not specify, and this is the one place in
the study where the difference shows. Second, on the predicted card the two predicates
separate widely --- 0.070 full match against 0.172 adherence --- and every row in that gap
is an artifact that executed the predicted card faithfully, which was itself the wrong
construction. The probe was run at fixed weights and has no stage in which the planner is
retrained, so it cannot say whether better planning would move the 0.070 toward the 0.210
the reference contract reaches.

\begin{table}[t]
\centering
\caption{The variational half of the 7B probe by card and predicate: the two rows answer
different questions and are not interchangeable.}
\label{tab:7b-vqe}
\footnotesize
\begin{tabular}{@{}lrrrr@{}}
\toprule
Card given & Full match & $n_{\mathrm{full}}$ & Adherence & $n_{\mathrm{adh}}$ \\
\midrule
predicted & 0.070 & 100 & 0.172 & 99 \\
reference & 0.210 & 100 & 0.210 & 100 \\
\bottomrule
\end{tabular}

\vspace{2pt}
{\footnotesize Adherence is decision-level (motif, repetition, sharing and tail); full match
adds the gate set and the arity and is scored against the reference construction.
$n_{\mathrm{full}}=100$ in both rows. $n_{\mathrm{adh}}=99$ on the predicted row because one
artifact could not be extracted; the reference row is out of 100 on both predicates.}
\end{table}

\emph{Answer to Question~4.} On the 3B conditions the card is carried out on QAOA
(adherence 0.808) and not on VQE (0.020). Full match moves only where it is carried
out, and the reference card takes the VQE parse rate from 0.810 to 0.550. The reading
and the data come from the same two families, so this is not prospective validation.
On the 7B probe the reference card reaches 0.210 on both predicates, while the
predicted card reaches adherence 0.172 and full match 0.070
(Table~\ref{tab:7b-vqe}): the model often carries out a wrong construction. A
two-stage, plan-free continuation at 7B is untested. The evidence fits a
generation-side failure under this field list, including the absent parameter
budget, and the design does not separate those reasons. The contract defined here is
insufficient for VQE on these conditions.

\subsection{Semantic measures after a shared refit}\label{sec:semantics}

On QAOA the structural gain moves some semantic summaries and not others. On the
primary measure, HQCR, the same-budget control is not separated. Every
difference is paired on rows where both conditions converged, so it need not equal the
gap between the rates in Table~\ref{tab:main}.

The shared refit converges on 94, 91, 87, 79, 93, 97 and 97 of 100 rows, in the order
of Table~\ref{tab:main}. The paired intersection is 90 on the two primary contrasts.
Non-converged rows are dropped, not imputed, so the HQCR claim is a complete-case
claim. Scoring non-convergence as failure could move the interval. The deposited logs
support that recompute. The structural contrasts use all 100 rows and do not depend on
it.

Against the same-budget control the distributional score moves by $+8.9\pt$, with
interval $[0.0,+17.8]$, $p=0.096$, $n=90$: the interval touches zero and the comparison is
not established at this $n$. Against the one-stage baseline it is $+17.6\pt$, interval
$[+7.7,+27.5]$, $p=0.0015$, $n=91$, but that baseline differs in the checkpoint as well
as in the prompt; that is why the same-budget row governs. The two-sided energy score
moves the same way and separates both: $+17.6\pt$ against one-stage on $16/0$ discordant
rows ($[+9.9,+25.3]$, $p=3.1\times10^{-5}$) and $+10.0\pt$ against matched-step on $9/0$
($[+4.4,+16.7]$, $p=0.0039$). The one-sided hit is thinner, $+6.7\pt$ on $6/0$
($p=0.031$). Against matched-step the distributional score moves least of the three.

What these numbers do not show is a movement of the typical artifact. On the converged
intersection the paired median difference in distance to the reference circuit's fitted
energy is 0.0000 between the reference and predicted contracts ($n=95$: 7 rows better, 8
worse) and 0.0000 between the predicted contract and the matched-step control ($n=85$: 23
better, 17 worse). Nine of those 85 rows crossing the 0.2 threshold is a change in the
tail, not in the median. On the measure this study designated primary the control is therefore not separated,
while two secondary scores move in the tail. The structural result is kept structural: the
gain is not credited with a semantic return in the abstract, the contributions or the
conclusion. The one-sided hit
is not the primary semantic measure, because against \goldc{} it equals 1.000 for both
conditions and cannot separate them; HQCR remains that measure.

A second refit with two restarts and 100 iterations, instead of five and 250,
depresses every arm. On QAOA the distributional difference between Conditioned and
Matched-step is $+16.7$ points there against $+8.9$ under the main protocol, and the
energy difference is $+12.2$ against $+10.0$ ($n=90$ in all four cells). Direction is
stable and significance is not. At the main budget the primary measure does not
separate the control ($p=0.096$). At the smaller budget it does ($+16.7$, 18/3
discordant, $[+7.8,+26.7]$, $p=0.0015$). That budget sits below what COBYLA needs for
the largest circuits, so the run is a stress test. The claim is the one the
main refit budget supports. VQE stays in the same low band under both protocols,
$-5.8$ against $-1.9$ points on the two-sided score.

\subsection{Overlap between training and evaluation}\label{sec:contam}

This benchmark is not instance-disjoint. In the 400-row test split from which the
evaluation set is drawn, 8.5\% of rows copy a training row verbatim and 74.5\% are
variants of a training instance. The 200 evaluation rows are a selected subset of that
split rather than a random draw (Section~\ref{sec:instances}), which is why 102 of them
are absent from every training split---far more than the parent split would contain.
Table~\ref{tab:strata} repeats the primary matched-step comparison on QAOA under four
overlap views and adds the pooled clean-strict cell over both families;
Table~\ref{tab:overlap} fixes the vocabulary, because a row can be an
instance variant and still carry a signature unseen in training.

The family-clean cell needs one qualifier before its $p$-value is read. It holds the 46
QAOA rows whose signature is absent from the QAOA and VQE training splits used here, and
the effect there is $+23.9\pt$ with $p=0.0074$. Those rows are not thereby unseen by every
model that produced the corpus's reference programs, since the corpus carries further
splits that this study neither used nor names. On the 20 rows absent from every split the
interval is $[0,+50]$; that interval, not the family-clean $p$-value, is the statement
about instances outside training.

\begin{table}[t]
\centering
\caption{The overlap vocabulary. ``Training row'' means a row of the corpus split
that was used for fine-tuning; the signature is the corpus identifier of an
instance, so two rows with the same signature are two renderings of one instance.}
\label{tab:overlap}
\footnotesize
\setlength{\tabcolsep}{3pt}
\begin{tabular}{@{}l >{\raggedright\arraybackslash}p{5.2cm}@{}}
\toprule
Term & Definition used here \\
\midrule
signature & the corpus identifier of an instance; equal signatures mean one
  instance rendered at different depths or qubit counts \\
verbatim duplicate & a test row whose prompt text and reference program both occur,
  character for character, in a training row \\
instance variant & a test row whose signature occurs in training but whose
  (family, qubits, depth) triple differs from any training row with that
  signature \\
family-clean & the signature occurs in no row of the QAOA or VQE training split \\
instance-disjoint & the signature occurs in no training row of any split \\
\bottomrule
\end{tabular}
\end{table}

\begin{figure}[t]
\centering
\includegraphics[width=\linewidth]{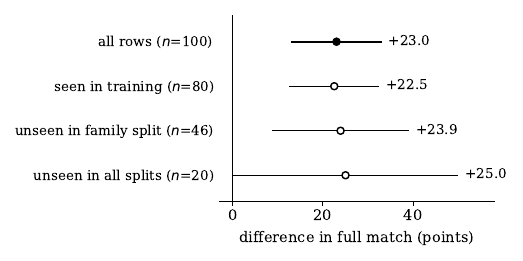}
\caption{The primary contrast on the QAOA overlap views of
Table~\ref{tab:strata}, drawn to the same scale. The interval widens as the view
becomes stricter because the number of rows falls; the point estimate does not
rise with instance exposure, and the two widest views cannot exclude a difference
of zero. The pooled clean-strict cell is tabulated but not drawn here.}
\label{fig:strata}
\end{figure}

Fig.~\ref{fig:strata} plots the same intervals. The views overlap: a signature absent from
the family training split can still occur elsewhere, so the counts 46 and 80 are not
complements of 100. What the four views show is narrower than the absence of memorization.
The seen view gives $+22.5\pt$ and the family-clean view $+23.9\pt$, so the estimated gain
does not rise with instance exposure. That pattern rules out the simplest verbatim-recall
account; a model that retained partial structure rather than whole instances could still
produce it, and this design would not separate the two.

Moderation is tested directly as the stratum effect minus the all-rows effect, with a
paired bootstrap over rows (20{,}000 resamples). On the family-clean stratum the difference
is $+0.9\pt$ with interval $[-10.0,+11.8]$; on the instance-disjoint stratum it is
$+2.0\pt$ with $[-22.8,+26.0]$. Any inflation from overlap is therefore bounded near twelve
points, and a deflation of about ten cannot be excluded. The second interval is wide enough
to contain the estimate itself; that is a power limitation
(Section~\ref{sec:th-conclusion}), not evidence that the effect shrinks. Pooling the
strictest definition over both families ($n=102$) yields $+5.9\pt$ $[0.0,+11.8]$,
$p=0.11$: that cell is not a substitute for the QAOA primary.

\begin{table*}[t]
\centering
\caption{The primary contrast by benchmark overlap: Conditioned against Matched-step on
full match. The first four rows are QAOA views from Table~\ref{tab:overlap}; the last
row pools the strictest definition over both families.}
\label{tab:strata}
\footnotesize
\begin{tabular}{@{}lrrrrr@{}}
\toprule
View & $n$ & Matched-step & Conditioned & Difference [interval] & $p$ \\
\midrule
All rows (QAOA)           & 100 & 0.550 & 0.780 & $+23.0$ $[+13.0,+33.0]$ & $3.4\times10^{-5}$ \\
Seen in training          &  80 & 0.600 & 0.825 & $+22.5$ $[+12.5,+32.5]$ & $1.2\times10^{-4}$ \\
Unseen in family split    &  46 & 0.500 & 0.739 & $+23.9$ $[+8.7,+39.1]$  & $0.0074$ \\
Unseen in all splits      &  20 & 0.350 & 0.600 & $+25.0$ $[0.0,+50.0]$   & $0.18$ \\
Pooled clean-strict (both families)
                          & 102 & ---   & ---   & $+5.9$ $[0.0,+11.8]$    & $0.11$ \\
\bottomrule
\end{tabular}

\vspace{2pt}
{\footnotesize Terms are defined in Table~\ref{tab:overlap}. The strict QAOA interval
is a power limitation (Section~\ref{sec:threats}). Cell rates are omitted from the
pooled row because the families are not pooled elsewhere. Difference of differences
against the all-rows estimate: family-clean $+0.9$ pt $[-10.0,+11.8]$;
instance-disjoint $+2.0$ pt $[-22.8,+26.0]$ (20{,}000 resamples).}
\end{table*}

\subsection{Where the fields fail}\label{sec:fields-fail}

Aggregate rates say how often a condition is right; they do not say which field decides a
miss. Table~\ref{tab:fields} breaks full match into the fields it conjoins. On QAOA,
allocation is satisfied by at least 0.88 of artifacts in every condition and is rarely the
deciding miss, while order equals cover in every arm because the reference order is the
cover order; both remain checks in the sense of Table~\ref{tab:fields-select}. The fields
that move with the conditions are the coefficient slot (0.550 to 0.880) and inversion
(0.670 to 0.940). Full match never exceeds the smaller of cover and slot. Renaming
disturbs the slot without changing the grouping (0.660); shuffling breaks the grouping
itself (0.610).

The VQE block shows the opposite pathology. Arity stays at or above 0.99, the repetition
count at or below 0.120, and the sharing pattern at or below 0.320. Motif and gate set
remain low in every condition, including \goldc{}, where the generator is handed the
reference contract. At field resolution the failing fields sit in the emitted artifact
even when the reference card is supplied, so planner field accuracy alone does not explain
the miss; the design still does not separate that pattern from the absent parameter-budget
field (Section~\ref{sec:vqe}).

If the fields failed independently, full match would be the product of its marginals. The
observed joint exceeds that product by a factor of 1.3 to 2.2 on QAOA and 10 to 40 on VQE.
The gap rules out the reading in which per-field slips are independent and could be
repaired one field at a time; it does not estimate which pairs move together.

\begin{table*}[t]
\centering
\caption{Field-level pass rates: the fraction of artifacts whose field agrees with the
reference construction, per condition and per family.}
\label{tab:fields}
\footnotesize
\setlength{\tabcolsep}{4pt}
\begin{tabular}{@{}lrrrrrr@{}}
\toprule
\multicolumn{7}{@{}l}{\emph{QAOA}: fields of the reference construction, 100 instances per condition} \\
Condition & cover & slot & alloc. & order & inv. & full match \\
\midrule
One-stage     & 0.780 & 0.570 & 0.950 & 0.780 & 0.790 & 0.510 \\
Matched-step & 0.850 & 0.630 & 0.970 & 0.850 & 0.770 & 0.550 \\
Equal-weight  & 0.740 & 0.550 & 0.950 & 0.740 & 0.720 & 0.460 \\
Shuffled      & 0.860 & 0.610 & 0.880 & 0.860 & 0.670 & 0.400 \\
Renamed       & 0.830 & 0.660 & 0.900 & 0.830 & 0.680 & 0.530 \\
Conditioned   & 0.920 & 0.840 & 0.950 & 0.920 & 0.900 & 0.780 \\
\goldc{}   & 0.930 & 0.880 & 0.950 & 0.930 & 0.940 & 0.870 \\
\midrule
\multicolumn{7}{@{}l}{\emph{VQE}: fields of the reference construction, 100 instances per condition} \\
Condition & motif & repeat & share & tail & gate set & full match \\
\midrule
One-stage     & 0.200 & 0.070 & 0.260 & 0.660 & 0.230 & 0.020 \\
Matched-step & 0.250 & 0.120 & 0.270 & 0.570 & 0.280 & 0.020 \\
Equal-weight  & 0.270 & 0.110 & 0.300 & 0.610 & 0.320 & 0.030 \\
Shuffled      & 0.430 & 0.090 & 0.300 & 0.540 & 0.500 & 0.040 \\
Renamed       & 0.270 & 0.110 & 0.300 & 0.590 & 0.320 & 0.030 \\
Conditioned   & 0.200 & 0.100 & 0.320 & 0.560 & 0.210 & 0.030 \\
\goldc{}   & 0.410 & 0.090 & 0.280 & 0.560 & 0.500 & 0.030 \\
\bottomrule
\end{tabular}

\vspace{2pt}
{\footnotesize Full match requires every field, so it lies below each column. Arity is
omitted: it is 1.000 except 0.990 on the two one-stage-trained conditions. Order equals
cover in every QAOA row. Allocation stays above 0.88 in every condition.}
\end{table*}

\subsection{Does structural match imply a better artifact?}\label{sec:construct}

A reference-based measure invites an obvious question: does it track anything the
consumer of the artifact cares about? The per-row data answer part of it for the two arms on which structural and semantic
outcomes both exist: Conditioned and Matched-step,
on QAOA, on the 188 rows whose shared refit converged. The first row of the table that
follows is definitional, and the reading below is built on the second.

\begin{table}[t]
\centering
\caption{Row-level association between structural match and semantic outcome, for
Conditioned and Matched-step on QAOA, over the rows on which the shared refit converged
($n=188$).}
\label{tab:assoc}
\footnotesize
\begin{tabular}{@{}lrrrr@{}}
\toprule
Group & $n$ & HQCR & $q$-hit@0.2 & SREV (2s) \\
\midrule
Full match     & 133 & 0.992 & 1.000 & 1.000 \\
No full match  &  55 & 0.400 & 0.891 & 0.764 \\
\bottomrule
\end{tabular}

\vspace{2pt}
{\footnotesize The first row is definitional and is a consistency check on the refit, not
evidence: an artifact whose construction matches can reproduce the reference distribution by
construction (Conditioned $0.987$, Matched-step $1.000$). The second row is the informative
one, and it holds in each arm separately (Conditioned $0.316$, Matched-step $0.444$). Values
are the fraction of rows on which the outcome holds. SREV
is the two-sided score of Section~\ref{sec:measures}: the fitted energy is within
0.2 of the reference energy in either direction. The one-sided hit rate is the
least informative of the three, holding on 89\% of the rows that miss the
construction, whereas agreement with the reference distribution falls to 40\%.}
\end{table}

\begin{table}[t]
\centering
\caption{What differs in the artifacts that miss the reference construction.
Conditioned and Matched-step, QAOA, rows on which the shared refit converged
($n=55$ mismatches out of 188). The first two rows change the construction; the
third changes only how the construction is annotated.}
\label{tab:mismatch}
\small
\setlength{\tabcolsep}{4pt}
\begin{tabular}{@{}lrrr@{}}
\toprule
What differs & $n$ & share & HQCR \\
\midrule
covered terms (cover)      & 16 & 29\% & 0.312 \\
grouping of symbols (slot) & 30 & 55\% & 0.400 \\
allocation, order or inversion only & 9 & 16\% & 0.556 \\
\bottomrule
\end{tabular}
\end{table}

The first row of Table~\ref{tab:assoc} is a consistency check on the shared refit, not a
finding: an artifact whose construction already matches can reproduce the reference
distribution once both circuits are fitted. The informative row is the second. Of the 55
artifacts that miss the construction, 0.891 still come within 0.2 of the reference energy
on the one-sided criterion, while agreement with the reference distribution falls to
0.400. That separation holds inside each arm as well (Conditioned 0.316 against 0.987;
Matched-step 0.444 against 1.000).

Table~\ref{tab:mismatch} answers the objection that a reference construction is only one of
several correct ones. Most mismatches are not of that kind: 84\% change either the covered
terms or the grouping of coefficients across symbols, which changes the parameterization
rather than the spelling. Of the remaining 16\%---allocation, order, or inversion
alone---only 0.556 reach the reference distribution, so the group is not a set of
equivalent artifacts that the measure wrongly rejects. A state-vector check on the same
188 rows (one uniform parameter draw per row, all-zero input, squared-overlap tolerance
$10^{-6}$) puts a number on that decomposition: all 30 comparable cover-or-slot mismatches
fall below the tolerance, three of the nine incidental mismatches do, and none of the 133
full matches is distinguished. An overlap below the tolerance refutes equality; an overlap
of one does not establish it. No consumer that reads the construction is run
(Section~\ref{sec:th-construct}).

\subsection{Is a legal contract enough?}\label{sec:relevance}

The controls so far either corrupt the reference contract or rename its symbols, and both
still share its vocabulary. Condition A1 asks whether any concrete plan in the prompt would
do: it hands the generator a shape-compatible but instance-irrelevant reference contract
from another instance of the same family, qubit count, and depth. The donor is a
deterministic function of the row key, so the condition does not depend on row order.

\begin{table}[t]
\centering
\caption{Relevance control on QAOA ($n=100$): the conditioned and the reference condition
against A1, which is given a shape-compatible but instance-irrelevant reference contract.}
\label{tab:relevance}
\footnotesize
\setlength{\tabcolsep}{4pt}
\begin{tabular}{@{}lrrr@{}}
\toprule
Comparison & A & B & Diff.\ [interval] \\
\midrule
A1 vs placeholder (\ew{})   & 0.130 & 0.460 & $-33.0$ $[-44,-22]$ \\
\ours{} vs A1               & 0.780 & 0.130 & $+65.0$ $[+55,+75]$ \\
\goldc{} vs A1              & 0.870 & 0.130 & $+74.0$ $[+65,+82]$ \\
\bottomrule
\end{tabular}

\vspace{2pt}
{\footnotesize A1 receives the reference contract of another instance of the same family,
qubit count and depth, written in the surface form of \goldc{}. A is the first condition of
the pair, B the second, and values are full-match rates; differences are paired, in percentage
points, with a row-level bootstrap interval. All three have exact McNemar $p<10^{-6}$.
Interface closure is reported beside the structural rates because a plan that does not fit the
instance also disturbs the declaration list, which is a second route to failure.}
\end{table}

An irrelevant but legal plan is worse than no plan at all: 0.130 against 0.460 for the
placeholder on the same weights ($-33.0\pt$). The gain therefore sits in the relevance of
the plan, not in its presence; giving these weights the right plan is worth $+32.0\pt$.
A1 also breaks interface closure more often (0.400 against 0.720). Restricted to the rows
on which both conditions close ($n=31$), A1 still reaches only 0.323 against 0.613
($-29.0\pt$), so the loss is not an artifact of the broken interface alone. Against the
donor contract rather than the reference, cover and slot agreement are 0 of 100: the
condition varies the applicability of a realized plan, not a uniform draw from the grammar
of legal contracts (Section~\ref{sec:th-controls}).

\section{Discussion}\label{sec:discussion}

\subsection{Implications for reporting and practice}\label{sec:disc-report}

The measurements support two reporting conventions. Interface success and structural
match should be separate columns: parse rates between 0.55 and 0.97 do not recover
full-match rates between 0.02 and 0.87. Where symbol names are arbitrary, skeleton
match should sit beside the strict predicate. The renaming control costs $34.0\pt$ on
full match and $17.0\pt$ on skeleton match. Contrasts should also carry the overlap
strata of Section~\ref{sec:contam}.

The supported beneficiary is a host that reads the construction. For that host,
Table~\ref{tab:cell} is the insertion rule: card content helps on weights trained with
cards and costs points on plan-free weights, and seed 1337 replicates the direction.
A host that thresholds fitted energy or a distributional score is a different
consumer. Against matched-step the primary HQCR contrast is not separated, and
the paired median energy gap is zero (Section~\ref{sec:semantics};
Table~\ref{tab:assoc}). For that consumer, the structural contrasts alone do not
justify inserting a card.

On this QAOA template the numbers support a shipping order. Fix the field list and the
checker. Prefer a rule when one recovers the card. Compile the card when a fixed gate
template exists, and call the generator when the card is not executable. On the 34
rows where the rule and the reference disagree, the generator follows the reference in
17 and the card in 4 (Section~\ref{sec:attrib}). Two stop rules follow. A card on
weights that never saw one costs points. On VQE the reference card cuts parse from
0.810 to 0.550 without buying structural match (Table~\ref{tab:vqe}). An adherence
pilot therefore precedes any deployment claim. A parameter-budget field is not shown
to repair that family. Transfer still needs a host-specific walk
(Section~\ref{sec:disc-transfer}).

\subsection{Cost}\label{sec:disc-cost}

The checker is linear in the number of gates and calls no simulator. Inference is two
forward passes. The shared refit costs about 36\,s on QAOA and 3.4--4.2\,min on VQE
(Table~\ref{tab:cost}). The dominant wall clock is the 638-step continuation
($\sim$103\,min), not the extra planner pass. Counted per matching QAOA artifact at
15\,s of generation, wall clock falls from 29\,s (one-stage) to 19\,s (conditioned)
because more rows match, not because each generation is cheaper.

\begin{table*}[t]
\centering
\caption{Cost of the two signals. The evaluator column describes the refit-based
score used for the semantic measures; the contract column describes the
intermediate artifact used at inference. Both sets of figures were measured on one
RTX 4090~D (24\,GB) with bfloat16 weights and greedy decoding
(Section~\ref{sec:instances}).}
\label{tab:cost}
\footnotesize
\begin{tabular}{@{}p{3.0cm}p{4.6cm}p{4.6cm}@{}}
\toprule
Quantity & Contract path & Evaluator path \\
\midrule
Payload & 76--1{,}489 chars & one scalar per artifact \\
In the context before generation & yes & no \\
Check & $O(\#\mathrm{gates})$ walk, no simulator & 36\,s (QAOA), 3.4--4.2\,min (VQE) \\
Check runs during decoding & no & no \\
Inference & two forward passes & one pass plus scoring \\
Locates the failing step & line level & no \\
\bottomrule
\end{tabular}

\vspace{2pt}
{\footnotesize The evaluator figures include the variational refit at the shared
budget. A single forward pass is common to both paths. Neither path replaces the
evaluator inside the decoding loop: the contract is written before generation and checked
after it, and no contrast here is trained against a contract reward.}
\end{table*}

\subsection{What transfers}\label{sec:disc-transfer}

The proposal is for embedded programs whose construction can be written as a short
vector of decisions and checked by a walk. It has been examined on OpenQASM~3 and on
two circuit families. Transfer to another DSL is a hypothesis, not a result. Another
language would need its own corpus, coverage count, and failure rates before a
contract there could be credited with a gain. VQE is the in-paper illustration:
without realization, full match stays between 0.02 and 0.04.

The field-list filters of Section~\ref{sec:fields} are a domain-general procedure,
not a result tested outside this corpus. Reference-based structural match is an
engineering property for a host that reads the
construction, not a universal correctness criterion. The three readings in
Section~\ref{sec:properties} were written after the fact. They separate these
families by adherence. They do not predict gain elsewhere, and they do not promise a
small field list on every corpus. Planner field accuracy is not a fourth reading
(Section~\ref{sec:attrib}). A hierarchy of contracts is unanalyzed. The VQE result
is consistent with a failing step below the contract.

\section{Threats to validity}\label{sec:threats}

\subsection{Construct validity}\label{sec:th-construct}

Full match records agreement with a reference construction. It does not record solution
quality, and it is not acceptance by a downstream tool: no such tool is run here. Most
mismatches differ as operators rather than as spelling
(Section~\ref{sec:construct}); of the nine that do not, four also miss the reference
distribution (Table~\ref{tab:mismatch}). Semantic measures bound the structural claim on QAOA only, where the shared
refit is closed and adherence is high (Section~\ref{sec:semantics}). Emitted-parameter
scores have median zero on every QAOA condition and are not used as outcomes.

\subsection{Internal validity}\label{sec:th-internal}

Symbol names have no meaning in the circuit semantics, yet a group-preserving renaming
still costs $17.0\pt$ on skeleton match. Both predicates are therefore reported. Refitting
is a possible confounder because conditions emit different numbers of parameters; the
budget is shared (five restarts, 250 iterations), and the VQE parameter-count gap is
stated explicitly. Renaming and shuffle plans are deterministic functions of the reference
contract and a fixed seed, recomputed by the evaluation scripts rather than stored as
separate logs.

\subsection{External validity}\label{sec:th-external}

Main generation uses one 3B checkpoint, greedy decoding, one planner, and one run per
condition, with two further training seeds for the primary pair; those seeds retrain the
continuation and hold the planner and the base checkpoint fixed. A 7B fixed-weight probe
(Section~\ref{sec:effect}) raises VQE adherence to 0.210 on the reference card and 0.172
on the predicted card, with predicted-card full match at 0.070
(Table~\ref{tab:7b-vqe}). It does not replicate the two-stage comparison, and decoding,
planner architecture, and pipeline are not varied. The contribution is a measurement
framework and a design proposal for this pipeline, not a claim that contract-conditioned
generation is universally effective. Two families are studied and are not pooled;
evaluation covers two of seventeen VQE ansatz families; the overlap strata are part of the
result rather than an afterthought. The QAOA evaluation set is dominated by the
$\le 8$-qubit band (75 of 100 rows), with 21 rows at $9$--$12$ qubits and 4 above 12; the
family-level full-match contrast is therefore a small-circuit estimate unless a later
reanalysis of the deposited logs shows otherwise.

\subsection{Conclusion validity}\label{sec:th-conclusion}

The instance-disjoint QAOA stratum has $n=20$. The point estimate is $+25.0\pt$ with
interval $[0,+50]$ and $p=0.18$. The study reads that cell as underpowered, not as evidence
of no effect, and does not replace it with either the QAOA all-rows estimate or the pooled
clean-strict cell of Table~\ref{tab:strata} ($n=102$, $+5.9\pt$ $[0.0,+11.8]$, $p=0.11$).
On VQE the interval $[-22.2,+4.4]$ excludes a gain of about 15 points and admits a
degradation of about 22; neither benefit nor harm is claimed. Overlap moderation is
$+0.9\pt$ $[-10.0,+11.8]$ (Section~\ref{sec:contam}). Semantic contrasts rest on the
refit-converged subset ($n=91$ and $90$). Whether the effect holds on instances the
generator has never seen is therefore not established here.

\subsection{The control that was not constructed}\label{sec:th-controls}

One comparison is absent and is recorded so that the controls that were run are not
expected to answer it: a contract drawn uniformly from the grammar of legal contracts,
rather than from realized contracts in the evaluation corpus. Renaming, shuffle, and A1
all remain close to realized plans. The hand-written rule of Section~\ref{sec:attrib} is
likewise not a row of Table~\ref{tab:main}: it reconstructs the reference contract on QAOA,
where consecutive equal-coefficient terms expose the grouping in the instance text, and
the finding that the planner is not what carries the effect is established on that family
only.

\section{Conclusion}\label{sec:conclusion}

A generated OpenQASM~3 fragment can pass its host and still miss the reference
construction. Across seven conditions, parse rates lie between 0.55 and 0.97 and full
match between 0.02 and 0.87. At the main refit budget the primary
distributional measure does not separate the same-step control, so the gain reported
here is structural (Section~\ref{sec:semantics}).

On the constructed QAOA set the card is worth $+32.0\pt$ at fixed card-trained
weights and costs $13.0\pt$ without that training. The training$\times$slot
interaction is $+45.0\pt$ and is not a deployment contrast. The same-step gain is
$+23.0\pt$. The one-stage gain is $+27.0\pt$ and also changes the checkpoint. A rule
card and a compiler of the predicted card are not separated from the learned pipeline.
Eighty of 100 signatures appear in some training split, and the 20-row
instance-disjoint interval includes zero.

On VQE, adherence is 0.020 and full match stays between 0.02 and 0.04. The reference
card leaves full match at 0.030 while cutting parse to 0.550, and the design does not
separate a missing parameter budget from other generation failures. The 7B probe
leaves that boundary open (Table~\ref{tab:7b-vqe}). For a host that reads the
construction, insert a card only into a generator trained to read it, prefer a rule
or a compiler when the template is fixed, and pilot adherence before any deployment
claim.

\section*{CRediT authorship contribution statement}

\noindent\textbf{Haowen Guo:} Conceptualization, Methodology, Software, Formal analysis,
Writing -- original draft.\\
\textbf{Yangzhao Huang:} Data curation, Validation,
Writing -- review \& editing.

\section*{Declaration of competing interest}

The authors declare that they have no known competing financial interests or
personal relationships that could have appeared to influence the work reported in
this paper.

\section*{Declaration of generative AI and AI-assisted technologies in the
manuscript preparation process}

During the preparation of this work the authors used Cursor to revise the
wording of the manuscript. After using this tool/service, the authors reviewed
and edited the content as needed and take full responsibility for the content of
the published article. The generative model evaluated in the study is named in
Section~\ref{sec:arms}; the assistance declared here was used for the manuscript,
not for producing the artifacts.

\appendix
\renewcommand{\thetable}{\Alph{section}.\arabic{table}}
\setcounter{table}{0}

\section{Contract grammar}\label{app:grammar}

A contract is a header followed by one decision line per step. The grammar is
small so that extraction and comparison are walks over the text.

{\footnotesize
\begin{verbatim}
contract ::= header line+
header   ::= "CARD" family
             "n=" int ["p=" int]
line     ::= index op set symbol
op       ::= "cost" | "lin" | "mixer"
             (* QAOA *)
set      ::= "[" int ("," int)* "]"
symbol   ::= "x" int
\end{verbatim}
}

\noindent On VQE the same header is followed by exactly five lines, because a
variational contract carries no symbol that a relabelling could act on.

{\footnotesize
\begin{verbatim}
vqe    ::= header motif repeat share
           tail gateset
motif  ::= "MOTIF rot=[" int ("," int)* "]"
           "ent=[" string ("," string)* "]"
           "full_pair=" ("0" | "1")
repeat ::= "R " int
share  ::= "SHARE " mode
mode   ::= "per_gate" | "per_qubit"
           | "shared"
tail   ::= "TAIL " ("rot1q" | "none")
gateset::= "GATESET " str ("," str)*
\end{verbatim}
}
\noindent Both families use the same header and the same line index, which is what makes
an index-wise comparison meaningful when the field predicates differ. The gateset and the
arity are checked once per artifact rather than per line.

\section{Prompt templates}\label{app:prompts}

Two templates are used, and both are frozen across conditions. The generator
receives the instance and a contract block. The planner receives the instance
only. The only text that changes between conditions is the contract block:
absent, a placeholder, a predicted contract, the reference contract, or a
corruption of the reference contract.

{\footnotesize
\begin{verbatim}
[system]  You are an expert at writing
          parameterized OpenQASM 3.0.
[user]    <instance text>
          [PLAN]
          <contract text>
[planner] <instance text>
          Emit the construction contract
          for this instance.
\end{verbatim}
}

\noindent The artifact repository contains the full strings, including the
placeholder used by Equal-weight and Matched-step. A one-token change in that
placeholder is a different condition, so the string is treated as data.

\section{Artifact layout}\label{app:artifact}

The deposit holds the extractor, the checker, both prompt templates, the per-row
generations, the evaluation scripts, and the script that rebuilds every table from
the logs. Table~\ref{tab:repro} maps each display to those files. A reader checking
one cell should start from the log of that condition: it carries the signature, the
emitted text, the extracted contract, and the three structural predicates.

\begin{table}[H]
\centering
\caption{Reproduction map. Scripts are in \texttt{code/part}; logs and tables are
in the artifact deposit. A row that lists two scripts reads the per-row log and
then aggregates it.}
\label{tab:repro}
\footnotesize
\setlength{\tabcolsep}{3pt}
\begin{tabular}{@{}l >{\raggedright\arraybackslash}p{3.5cm} >{\raggedright\arraybackslash}p{3.2cm}@{}}
\toprule
Output & Computed from & Script \\
\midrule
Tables~\ref{tab:main}--\ref{tab:vqe} & \texttt{arms/*.jsonl} &
  \texttt{part\_arms\_eval.py}, \texttt{\_struct\_fields.py} \\
Table~\ref{tab:paired} & \texttt{arms/*\_rows.jsonl} &
  \texttt{seed\_replication.py} \\
Table~\ref{tab:strata} & \texttt{keys200.jsonl}, \texttt{arms/*\_rows.jsonl} &
  \texttt{strat\_homogeneity.py} \\
Table~\ref{tab:fields} & \texttt{STRUCT\_FIELDS.json} &
  \texttt{\_struct\_fields.py} \\
Table~\ref{tab:assoc} & \texttt{arms/*\_rows.jsonl}, \texttt{REFIT\_V2\_rows.jsonl} &
  \texttt{row\_association.py} \\
Table~\ref{tab:relevance} & \texttt{arms/A1\_rows.jsonl} &
  \texttt{seed\_replication.py} \\
Table~\ref{tab:cost} & \texttt{REFIT\_V2.json}, run logs &
  \texttt{part\_refit\_v2.py} \\
Figs.~\ref{fig:contrasts} and~\ref{fig:strata} & Tables~\ref{tab:paired}
  and~\ref{tab:strata} & \texttt{make\_figures.py} \\
Fig.~\ref{fig:example} & \texttt{arms/M1.jsonl}, \texttt{arms/P0.jsonl},
  \texttt{gold\_cards\_test.jsonl} & none (excerpts printed from the logs) \\
\bottomrule
\end{tabular}
\end{table}

\section*{Data availability}

The extractor, checker, prompt templates, per-row generations, evaluation scripts,
and per-row logs are openly available as version v1 of \citet{guo2026artifact}
(\url{https://doi.org/10.5281/zenodo.23885644}). From those logs a third party can
rebuild every numbered table (Table~\ref{tab:repro}). The circuit corpus used for
evaluation and for deriving the contract has no public identifier. Its license
permits redistribution of derived artifacts but not of the raw circuits, which are
not in the deposit. Reference constructions therefore cannot be regenerated from a
public problem set. Derived indices and reference cards are included.

\section*{Funding}

This research did not receive any specific grant from funding agencies in the
public, commercial, or not-for-profit sectors.

\section*{Acknowledgments}

The authors thank colleagues who read earlier drafts for insisting that a structural
predicate and a semantic score be reported separately.
Section~\ref{sec:measures} is organized around that distinction.

\bibliographystyle{elsarticle-harv}
\bibliography{name}

\end{document}